\documentclass[aps,prl,reprint,superscriptaddress]{revtex4-2}
\usepackage{overpic}
\usepackage{amsmath,amssymb}
\usepackage{graphicx}
\usepackage{lipsum}  
\usepackage{hyperref}
\newcommand{\mean}[1]{\left\langle #1 \right\rangle}
\newcommand{\bk}{{\boldsymbol{k}}}
\newcommand{\de}{\partial}

\begin{document}

\title{The off-equilibrium Kinetic Ising model: The Metric Case}

\author{Luca Di Carlo}
\affiliation{Joseph Henry Laboratories, Princeton University, Princeton, NJ 08544, USA}
\affiliation{Lewis-Sigler Institute for Integrative Genomics}

\begin{abstract}
    We investigate the critical behavior of the Kinetic Ising model with non-reciprocal nearest neighbors interactions. A finite-size scaling study suggests that the model belongs to the Ising universality class.  We characterize the off-equilibrium behavior of the model by measuring the entropy production rate and by studying the coarsening dynamics, which shows a super-diffusive relaxation. A coarse-grained equation of motion is in compelling agreement with the numerical findings. A one-loop renormalization group calculation shows that the model belongs to the Model A universality class.
\end{abstract}

\maketitle
Universality is a fundamental concept in the study of critical phenomena and collective behavior. Universality refers to the observation that different systems can exhibit the same collective behavior near their critical points. This means that diverse systems, despite having distinct microscopic interactions and details, share identical critical behavior characterized by the same critical exponents and scaling laws \cite{kardar2007statistical}.
This allows for a unified understanding and a classification of critical phenomena across various physical contexts, thereby highlighting the underlying simplicity and coherence in the complexity of phase transitions.
The renormalization group (RG) theory provides a robust framework to understand this scale invariance and universality at critical points.

Extending the theory of critical phenomena to the realm of off-equilibrium systems presents a hard and yet necessary challenge. Nonequilibrium systems, such as those found in biological contexts, constantly operate far from thermodynamic equilibrium, and are characterized by constant heat dissipation and entropy production. Investigating critical phenomena in this context could provide deeper insights into the fundamental principles of self-organization, pattern formation, and robustness in living systems \cite{Honeberg93, marchetti2013hydrodynamics, fruchart2021non, Kuramoto2018, rouches2024polymer}.

In recent decades several minimal models of nonequilibrium collective dynamics have been proposed \cite{Vicsek95, Solon15, caballero2018strong, vansaders2023informational}. In this work, we study a simple off-equilibrium generalization of the Ising model, the Kinetic Ising model \cite{aguilera2021unifying}. In constrast to the traditional Ising model  which is defined by its steady-state Boltzmann probability distribution
\begin{equation}
    P( \boldsymbol{\sigma}) = \frac 1 Z e^ {\sum_{\rm ij} J_{\rm ij} \sigma_{\rm i} \sigma_{\rm j}}
    \label{eq:P_ising}
\end{equation}
the Kinetic Ising model is defined by the update probability, 
\begin{eqnarray}
    P(\boldsymbol{\sigma} _{t+1}| \boldsymbol{\sigma}_t)  = \prod _{\rm i} \frac{ e^{ h_{\rm i} ^t \sigma_{\rm i}^{t+1}}}{ 2 \cosh \left (h_{\rm i}^t \right )}
    \label{eq:K_Ising} 
\end{eqnarray}
where the local field is defined as $h_{\rm i}^t = \sum_{\rm j} J_{\rm ij} \sigma_{\rm j}^t$. It is worth noting the similarity between Eq.~\eqref{eq:K_Ising} and the Markov Chain Monte Carlo update rule used to sample the Ising probability distribution~\eqref{eq:P_ising}. The key difference between~\eqref{eq:P_ising} and ~\eqref{eq:K_Ising} lies in the symmetry properties of the interaction matrix $J_{\rm ij}$. While in~\eqref{eq:P_ising} the interactions $J_{\rm ij}$ must be symmetric - or more precisely only the symmetric part of $J_{\rm ij}$ is relevant - in the kinetic Ising model there are no constraints on the symmetry of the interactions $J_{\rm i j}$. If $J_{\rm ij}$ is symmetric the Markov chain defined by~\eqref{eq:K_Ising} reproduces the Ising equilibrium probability distribution~\eqref{eq:P_ising}. The fully connected Kinetic Ising model has been extensively studied in the literature \cite{aguilera2021unifying,Kappen00}. In this work we study the asymmetric Ising model with metric interactions; more specifically we focus on a two-dimensional Ising model with non-reciprocal nearest neighbor interactions.  We show that, albeit being off-equilibrium, the system belongs to the same static universality class of the equilibrium Ising model. We characterize the off-equilibrium behavior of the system by computing the steady-state entropy production rate and by studying the coarsening dynamics, which is significantly affected by the non-reciprocal interactions. Finally, we study the corresponding coarse-grained field theory, which is in compelling agreement with the numerical results.  
\section{The Asymmetric Metric Ising model}
\begin{figure*}
    \includegraphics[width = \textwidth]{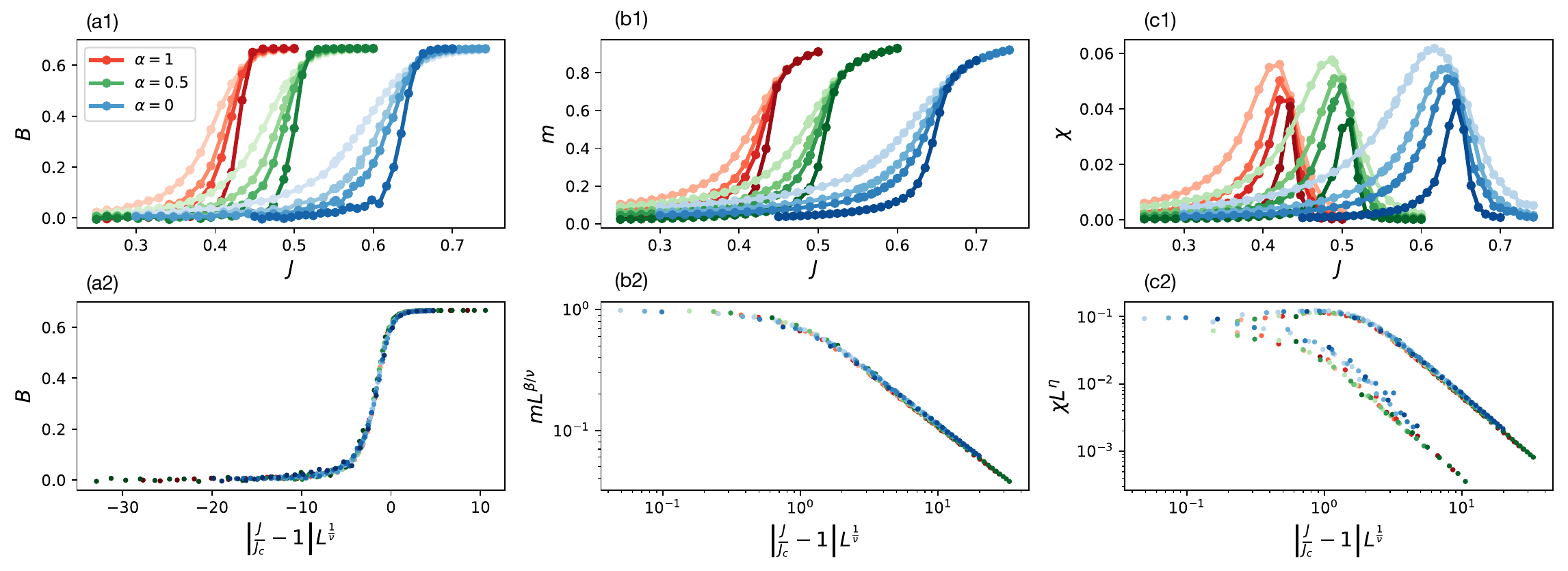}
    \caption{Finite-size scaling analysis. {\bf(a1)} The binder cumulant \cite{binder2005finite} as a function of the coupling $J$ for different values of the off-equilibrium parameter $\alpha$. Different colors corresponds to different values of $\alpha$, the color gradient indicates the size of the system ranging from $L = 16$ to $L = 64$.  The critical value of the coupling $J_c$ depends on the value of $\alpha$. It shifts from its equilibrium value ($\alpha = 1$) $J_c \simeq 0.442$  to $J_c \simeq 0.7$ for $\alpha = 0$. {\bf(a2)} Collapse of the binder cumulant with $\nu = 1.0$. \textbf{(b1, b2)} Data collapse of the magnetization with the 2dimensional Ising critical exponents $\beta  =0.125$.  \textbf{(c1, c2)} Collapse of the susceptibility with the two-dimensional Ising critical exponent $\eta = 0.25$. While the value of the critical coupling $J_c$ depends on $\alpha$, the measure critical exponents and the scaling functions seems to be the same as the equilibrium Ising model independently on how far the model is from equilibrium. The scaling of the susceptibility in the ordered phase seems to show some dependence on the value of $\alpha$.  }
    \label{fig:scaling_panel} 
\end{figure*}
We consider Ising spins $\sigma_{\rm ij} = \pm$ on a two-dimensional lattice, interacting through a symmetric interaction along the y-axis and an asymmetric interaction along the x-axis:  
\begin{equation}
    \begin{split}
        J_{(\rm i_x, i_y), (\rm i_x, i_y) } = J &  \delta_{\rm i_x, j_x} \left[ \delta_{\rm i_y , {j_y+1}} +  
        \delta_{\rm i_y , {j_y-1}}\right] \\   + &  J  \delta_{\rm i_y, j_y }\left[  \delta_{\rm i_x, j_x-1} + \alpha \delta_{\rm i_x, j_{x}+1} \right] 
    \end{split}
\end{equation}
where the parameter $\alpha$ controls the strength of the asymmetry. We call this model the Asymmetric Metric Ising (AMI) model.  For $\alpha = 1$ we recover the equilibrium two-dimensional Ising model, while for any $\alpha \neq 1$ the system is out of thermodynamic equilibrium.  Similarly to what happens in the equilibrium Ising model, we expect a phase transition from disorder to order as we increase the interaction strength $J$. The level of asymmetry plays an important role, in fact if the asymmetry is too strong ($\alpha<0$) the system might present dynamical instabilities. For instance, in a system made of just  two spins interacting with $J_{\rm{12}} = - J_{\rm{21}}$,  the spin  $\sigma_{\rm 1}$ wants to point in the same direction as $\sigma_{\rm 2}$, while $\sigma_{\rm 2}$ wants to point in the opposite direction of $\sigma_{\rm 1}$; this leads to an oscillatory behavior. We expect the phenomenology of the $\alpha <0$ phase to be very rich. In this work we focus on the ferromagnetic part of the phase diagram, corresponding to $\alpha \geq 0$, and leave the $\alpha <0$ regime to future studies. In the $\alpha \geq 0$  regime we expect the system to exhibit a ferromagnetic phase transition.  

\begin{figure}
	\includegraphics[width = 1.0\linewidth]{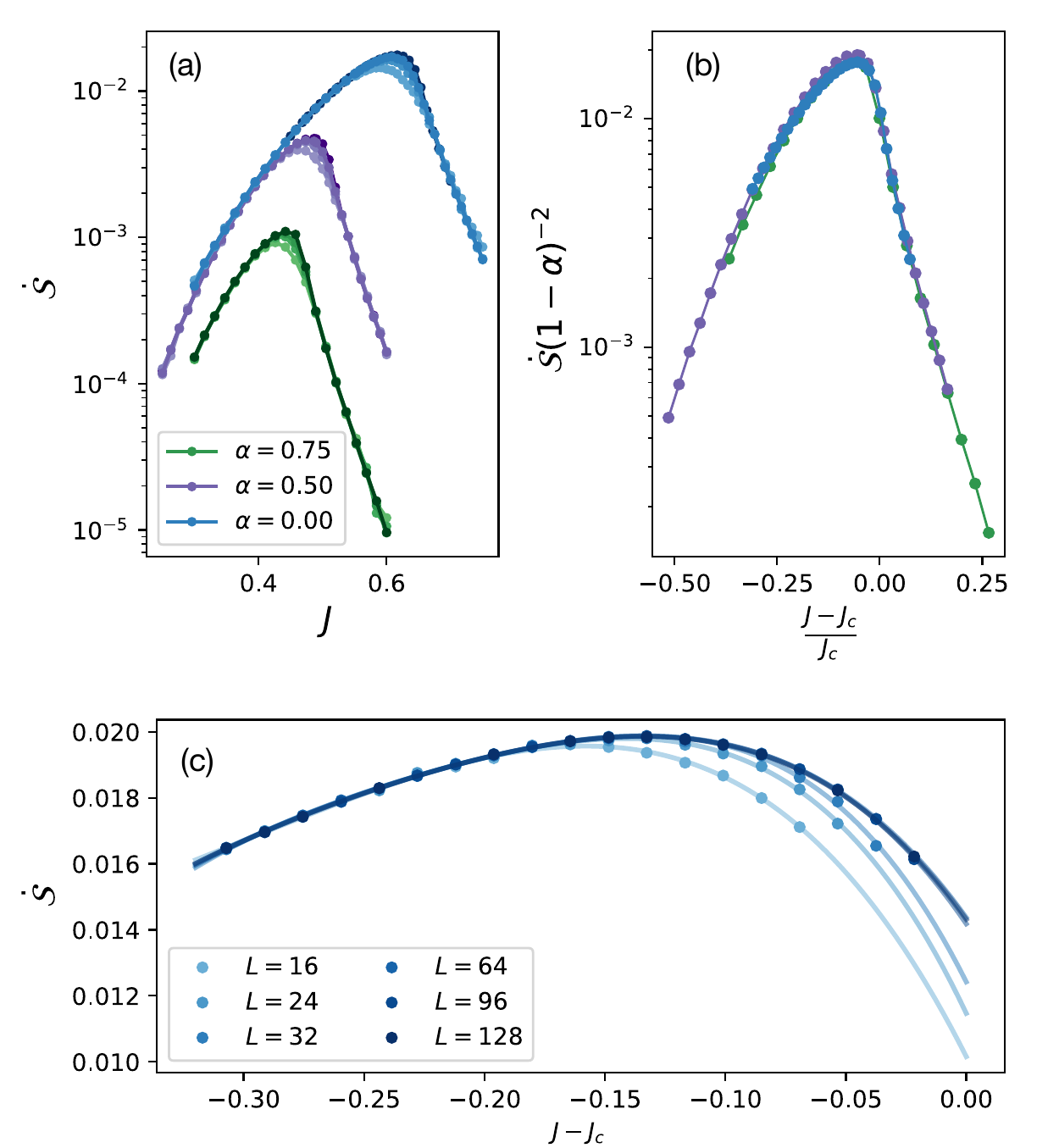}
    \caption{ Entropy production rate. \textbf{(a) - }  The entropy production rate of the model peaks around the transition point, unlike what observed in the other moles the entropy production rate is not cusped at the transition point. The color gradient indicates the system's size $L = 16, 24, 32, 64$. \textbf{(b) - } The entropy production rate seems to be well described by the ansatz~\eqref{eq:Sansa}, suggesting that $\dot{\mathcal S}$ is a simple quadratic function of the off-equilibrium parameter $(1-\alpha)$. \textbf{(c) -} Peak of the entropy production for increasing values of the system size $L$. The curve $\dot { \mathcal S}(J)$ seems converge to a smooth function.  }
    \label{fig:entropyProduction}
\end{figure}

We first address the effect of the asymmetry on the static critical properties of the model - namely we consider only observables measured at the same time. As expected, above a critical coupling $J_c$ the system exhibits a phase transition from a disordered to an ordered phase. As show in Fig.~\ref{fig:scaling_panel} (a1, a2, a3), the critical value of the coupling $J_c$ depends on the strength of asymmetry, going from its equilibrium value ($\alpha = 1$) $J_c \simeq 0.442$  to $J_c \simeq 0.7$ for $\alpha = 0$. A finite-size scaling analysis shows that the static critical exponents and the scaling functions are the same as the equilibrium Ising model, suggesting that the asymmetry does not affect the static universality class of the model. 
Although the equal time behavior is not affected by the asymmetry, this does not mean that the model is at equilibrium. The system's steady state violates both parity and time-reversal symmetries, as it is evident by inspecting the systems's trajectories~\footnote{ See trajectory.gif in the supplemental material}. If we interpret the spins as the occupation of a lattice site by a particle, the asymmetry generates a constant flux of particles in the direction of the asymmetry. We can quantify this flux using the following quantity, 
\begin{equation}
    \Phi  = \frac 1 {L^2} \sum_{\rm{x,y}}\mean{ \sigma_{\rm x+1,y}(t) \sigma_{\rm x,y}(t+\Delta t)} - \mean{ \sigma_{\rm x-1,y}(t) \sigma_{\rm x,y}(t+\Delta t)}
    \label{eq:PhiDef}
\end{equation}
which measures the average difference in probability between hopping to the right and hopping to the left. 
  Interestingly, $\Phi$  coincides, up to a numerical factor, with the entropy production rate of the system \cite{Crooks99, seifert2012stochastic, nardini17}. The per-particle entropy production rate can be written as \cite{aguilera2021unifying}, 
\begin{equation}
    \dot{ \mathcal {S}} = \frac 1 {L^2} \sum_{\rm ij} J_{\rm ij} \left( \mean{\sigma_{\rm i}^t \sigma_{\rm j}^{t-1}} -  \mean{\sigma_{\rm j}^t \sigma_{\rm i}^{t-1}} \right) = ( 1 - \alpha) \Phi 
\end{equation}
This equation has a clear interpretation. The energy that is dissipated by the system to  maintain the off-equilibrium steady-state is the product of the "net force" on the $x$ axis $(1-\alpha)$  and the 'velocity' $\Phi$ along the same direction. We expect the flux $\Phi$ to vanish at equilibrium $\alpha = 1$, and at the leading order we have $\Phi \simeq f(J, L) (1-\alpha)$.  Therefore the entropy production is quadratic in the deviation from equilibrium $(1-\alpha)$, 
\begin{eqnarray}
    \dot { \mathcal S } = f(J,L) (1- \alpha)^2
    \label{eq:Sansa}
\end{eqnarray}
ensuring that $\dot { \mathcal{S}}$ is positive.  In Fig.~\ref{fig:entropyProduction} (b,c) we show the entropy production rate for different values of  $\alpha$. In agreement  with the previous literature the entropy production peaks around the transition point \cite{Yu22,Ferretti22}. In particular, $\dot { \mathcal S}$ is maximum for $J$ slightly below the critical coupling $J_c$; this does not seem to be a finite-size effect, as shown in Fig.~\ref{fig:entropyProduction} (c). Fig.~\ref{fig:entropyProduction}(c) shows that the ansatz~\eqref{eq:Sansa} well describes what is observed in the numerical simulations. As opposed to the entropy production rate of other off-equilibrium model, such as the Vicsek model~\cite{Ferretti22} or the Active Ising model~\cite{Yu22}, which is cusped at the transition point,  we observer a smooth entropy production for $J \simeq J_c$, as shown in Fig.~\ref{fig:entropyProduction} (c). 

\begin{figure*}
    \includegraphics[width = 1\textwidth]{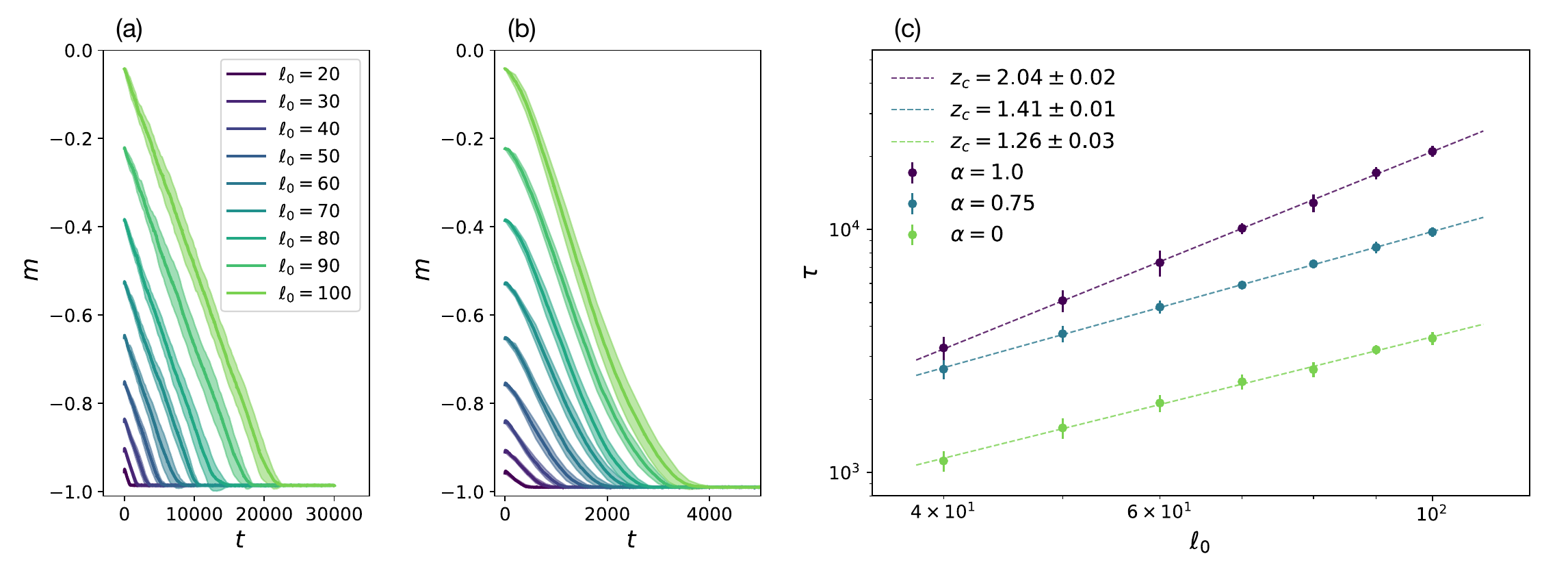}
    \caption{ Study of the coarsening dynamics \textbf{(a) - } Coarsening dynamics of the system in the ordered phase, $J = \frac 32 J_c$ , for $\alpha = 1$. For $\alpha = 1$ the system is at equilibrium and the coarsening dynamics is diffusive. Since we are considering a two-dimensional system, the magnetization relaxes linearly to its steady state value. Results are averaged over 20 different trials, error bars have been obtained as the standard deviation over the trials \textbf{(b) - } Coarsening dynamics in the ordered phase, $J = \frac 32 J_c$, for $\alpha = 0$, the statistics of equal time observable of this model matches the one of the $\alpha = 1.0$ system showed in panel(a). The coarsening dynamics of the off-equilibrium model ($\alpha = 0$) is significantly faster (see the coarsening.gif file in the SM). The magnetization is not decreasing linearly, suggesting a super-diffusive behavior.  \textbf{(c) - }   Scaling of the time it takes for the system to system to relax back to the steady state when the system is initialized in a configuration in which all the spin inside a ball of radius $\ell_0$ are pointing up, and all the spins outside the ball are pointing down. Data are relative to $J = \frac 32 J_c$. As we can see the relaxation time follows a scaling relation $\tau \sim \ell _0 ^{Z_{\rm c}}$, where the exponent $z_{\rm c}$ goes from $z_{\rm c} = 2$ at equilibrium  to lower and lower values as we move away from  equilibrium. Data are relative to a system of size $L = 256$. }
    \label{fig:coarsening} 
\end{figure*}

An other macroscopic consequence of the non reciprocity of the interactions regards the coarsening dynamics, which is the process of relaxation to equilibrium through the growth of domains of aligned spins. To study the coarsening dynamics we consider a system in the low temperature phase, initialized in a configuration in which all the spin inside a ball of radius $\ell_0$ are pointing up, and all the spins outside the ball are pointing down. We study the coarsening dynamics by measuring how long does it take for the system to relax back to $m \simeq -1$. According to the equilibrium theory of ordering kinetics \cite{bray2002theory} the positively magnetized region shrinks linearly in time, $\ell(t) = \sqrt{\ell_0 ^2 - A t}$, implying that the characteristic time scales to relax to the equilibrium state $\tau$ grows quadratically with the size of the magnetized region $\tau \sim \ell_0^2$.  Essentially, at equilibrium the relaxation process is purely diffusive. In Fig.~\ref{fig:coarsening} (a) we show the coarsening dynamics of the system in the ordered phase, $J = \frac 32 J_c$ , for $\alpha = 1$. On the contrary, When the system is off-equilibrium  $\alpha = 0$, the relaxation dynamics is significantly faster  ( see \verb|coarsening.gif| in the supplemental material), and the magnetization does not decrease linearly in time, suggesting a super-diffusive behavior. In Fig.~\ref{fig:coarsening} (c) we show the scaling of the relaxation time as a function of the size of the magnetized region $\ell_0$. The relaxation time follows a scaling relation $\tau \sim \ell _0 ^{z_{\rm c}}$, with a coarsening exponent $z_{\rm c}$  that goes from $z_{\rm c} = 2$ at equilibrium, in agreement with the theory of ordering kinetics,  to lower and lower values as we move away from the equilibrium regime, As we move away from $\alpha = 1$, the relaxation dynamics seems to approach a balistic regime $z_{\rm c} = 1$. 

\section{Coarse-grained Field theory}
The coarse-grained equation of motion of the AMI  can be derived by symmetry arguments. The minimal coarse-grained model compatible with the symmetries of the model takes the following form,  
\begin{eqnarray}
    \de_t \varphi = \Gamma_0 \nabla ^2 \varphi + v_0 \de_x \varphi - (m_0 + \lambda_0 \varphi^2) \varphi + \xi 
    \label{eq:EOM}
\end{eqnarray}
where $\xi(x,t)$ is a delta-correlated Gaussian noise, $\mean{ \xi(x,t) \xi(x^\prime, t^\prime) } = 2 \tilde \Gamma_0 \delta(x - x^\prime) \delta( t - t^\prime)$. The term $v_0 \de_x \varphi$ stems for the asymmetric interaction along the $x$ direction. When $v_0 = 0$ the equation of motion~\eqref{eq:EOM} coincides with Model A\cite{hohenberg1977theory}, which describes the relaxation dynamics of the equilibrium Ising model. In the Gaussian approximation, the correlation function of the AMI is given by,  
\begin{equation}
C_0(k, \omega) = \frac{2 \tilde \Gamma_0 }{ (\omega + v_0 k_x) ^2 + ( \Gamma_0  k^2 + m_0)^2} \qquad, 
\label{eq:GaussianC}
\end{equation}
and the static Gaussian correlation function $C_0(\bk, t= 0$) is, 
\begin{eqnarray}
    C_0(\boldsymbol{k}, t = 0) = \frac{\tilde \Gamma_0}{\Gamma_0 k^2 + m_0 } \qquad.
    \label{eq:Cstatic}
\end{eqnarray}
This suggests that the static properties of the model are independent of the value of the macroscopic parameter $v_0$. This results relies only on the fact that the off-equilibrium parameter $v_0$ appears in the combination $\omega + v_0 k_x$; when computing the Fourier transform of  \eqref{eq:GaussianC} at $t=0$ we can redefine the integration variable as  $\omega =  \omega^\prime - v_0 k_x $ to  reabsorb the term proportional to $v_0$. Since this results relies only on the form of the Correlation function, we expect it to hold also at the non-Gaussian level. 
A simple renormalization group calculation shows that the static critical properties are described by the equilibrium model A critical point ( see SM) \cite{hohenberg1977theory}.

\begin{figure}
    \includegraphics[width = 1.0\linewidth]{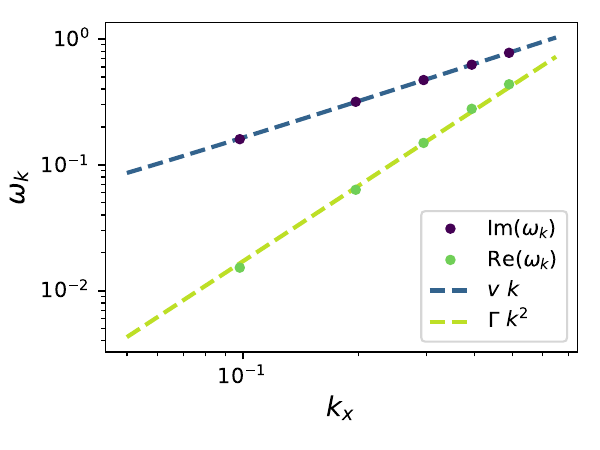}
    \caption{ Real and imaginary part of the characteristic frequency $\omega_{\boldsymbol{ k}}$ measured in the numerical simulations. The frequency is obtained by fitting Eqs.~\eqref{eq:Ckt} and~\eqref{eq:GaussianOmegaK}. The scaling of the real and imaginary parts of $\omega_{\boldsymbol k}$ are  in agreement with the Gaussian theory,  $\mathrm{Re}(\omega_{\bk}) \sim k^2$ and $\mathrm{Im}(\omega_{\bk}) \sim k$, as predicted by the renormalization group calculation (see SM). Data are relative to a system with $\alpha = 0.5$, $ J = 0.99 J_c$ of size $L = 64$. More details on how the characteristic frequency is measured can be found in the SM. }
    \label{fig:correlations} 
\end{figure}

The presence of the term $v_0 \de_x \varphi$ in the correlation function~\eqref{eq:GaussianC} suggests a non exponential relaxation; the Gaussian correlation function in the $(\bk, t)$ domain reads, 

\begin{eqnarray}
    C_0(\boldsymbol{k}, t)  =  \mathrm{Re} \left ( \frac{\tilde \Gamma_0}{\Gamma_0 \bk^2 + m_0} e^{ + i \omega_{\boldsymbol{k}}t}   \right) 
    \label{eq:Ckt}
\end{eqnarray}
where the frequency $\omega_{\boldsymbol k}$ is defined as, 
\begin{equation}
    \omega_{\boldsymbol k} = i \left( m_0 + \Gamma \boldsymbol{k}^2 \right) + v_0 \boldsymbol{k}_x 
    \label{eq:GaussianOmegaK}
\end{equation}

In Fig.~\ref{fig:correlations} we show the real and imaginary part of the characteristic frequency measured in numerical simulations close to the critical . In agreement with a renormalization group calculation (see SM), close to the critical point $\omega_{\bk}$ is well described by the Gaussian theory $\mathrm{Re}(\omega_{\bk}) \sim k^2$ and $\mathrm{Im}(\omega_{\bk}) \sim k$.

We finally compute the intensive entropy production rate in the Gaussian approximation. The entropy production rate can be computed by using the following Harada-Sasa relation \cite{Harada2005,nardini2017}, relating the entropy production rate to the violation of the fluctuation dissipation relations, 
\begin{equation}
    \mathcal S(k, \omega) = \frac {1}{ L^d} \frac{ L^d  \omega}{(2 \pi)^{d+1} \tilde \Gamma_0} \left[  \omega C(\bk, \omega) - 2 \tilde \Gamma_0 \mathrm{Im} \left ( R(\bk, \omega) \right) \right] 
    \label{eq:Sdot0}
\end{equation}
At the Gaussian level this formula leads to, 
\begin{equation}
    \dot {\mathcal {S}}( \bk ) =  \frac { v_0^2 k_x^2 }{ (2 \pi)^{d}} \frac{1}{ \Gamma_0 k^2 + m_0}
\end{equation}
This simple calculation is in agreement with the the numerical results shown in Fig.~\ref{fig:entropyProduction}. The entropy production rate is a quadratic function of the off-equilibrium parameter $v_0 \simeq (1-\alpha)$, and it peaks around the transition point $m_0 = 0 $. The $k_x^2$ factor in the numerator ensures that the entropy production rate $\dot { \mathcal S} ( \bk  \to 0 )$ remains finite also at the critical   point $m_0=0$. The $L^d$ factors in Eq.~\eqref{eq:Sdot0} cancel out and the intensive/per-particle entropy production rate is independent of the system size. 
\section{Discussion}
We have introduced the Asymmetric Metric Ising model,  a two-dimensional Ising model with non-reciprocal interactions, and we have studied its critical properties. As its equilibrium counterpart the system exhibits a ferromagnetic phase transition.  A numerical finite-size study shows that the model belongs to the Ising model (Model A) universality class, while being off-equilibrium. As observed in similar contexts \cite{caballero20, caballero2018bulk}, belonging to an equilibrium static universality class does not necessarily imply that the model is at equilibrium; In fact, the steady state entropy production rate of the model is non zero and it peaks close to the transition point.  Unlike observed in other contexts, the entropy production rate is smooth close to the transition point, as opposed to what observed in other off-equilibrium cases.  We characterized the coarsening dynamics in the low temperature phase, observing a significant speed up due to the non-reciprocal interactions.  In particular, we observe a super-diffusive relaxation to steady state. 
We finally studied the coarse-grained equation of motion of the model, and showed that at the one-loop level the model belongs to the Model A universality class. The off-equilibrium parameter $v_0$ does not take any renormalization group corrections, correctly predicting the scaling of the characteristic frequency measured in the numerical simulations. We computed the entropy production rate of the model using a generalized Harada-Sasa relation.  The numerical results are in excellent agreement with the theoretical predictions. This worked showed how non trivial behavior can emerge when combining the non-reciprocal interactions o with a metric structure, we expect that the strongly asymmetric regime $\alpha <0$ to have an even richer phenomenology. We are confident that the study of this simple system might give insights on the behavior of more complicated off-equilibrium phase transition. We believe that the phenomenology of the AMI will find application in describing real biological system, in particular the super-diffusive  relaxation dynamics  and the emergent collective transport.

\begin{acknowledgments}
    This work was supported by the National Science Foundation, through the Center for the Physics of Biological Function (PHY–1734030). I would like to thank Q. Yu for comments on the manuscript and M. Scandolo and C.W. Lynn  for useful discussions.  
\end{acknowledgments}

\bibliography{references}
\end{document}


\title{The off-equilibrium Kinetic Ising model : the metric case \\ 
supplemental material }

\author{Luca Di Carlo}
\affiliation{Joseph Henry Laboratories, Princeton University, Princeton, NJ 08544, USA}
\affiliation{Lewis-Sigler Institute for Integrative Genomics}

\maketitle

This supplemental material provides additional details and clarifications for the results presented in the main text of the paper. 

\section{Renormalization group calculation} 
In the main text we anticipated that it is possible to show that at one loop level the metric off-equilibrium Ising model belongs to the Ising universality class. To show this we need to compute the renormalization group flow of the parameters of the model. Following the standard procedure, we convert the stochastic differential equation, 
\begin{eqnarray}
    \de_t \varphi = \Gamma_0 \nabla^2 \varphi - m_0 \varphi - J \varphi^3  +v_0 \de_x \varphi + \xi 
    \label{mamal}\\ 
    \mean{\xi(x,t) \xi(x^\prime, t^\prime)} = 2 \tilde \Gamma_0 \delta(x-x^\prime) \delta(t-t^\prime)
\end{eqnarray}
into the following Martin-Siggia-Rose action\cite{tauber2014critical}, 
\begin{equation}
    S[ \varphi, \hat \varphi ] = \int d^d x dt \left[ \hat \varphi \left( \de_t  -  v_0 \de_x  - \Gamma_0 \nabla^2  - m_0   \right)\varphi   - J_0 \hat \varphi \varphi^3 + 2 \tilde \Gamma_0 \hat \varphi^2 \right]
\end{equation}
At the Gaussian level, the response $G_0(p, \omega) $ and correlation function $C_0(p, \omega)$ in Fourier space are given by,
\begin{eqnarray}
    G_0(\boldsymbol{p}, \omega) &=& \frac{1}{ - i \omega - i v_0 p_x  + \Gamma_0 \boldsymbol{p}^2 + m_0} \qquad \qquad \qquad  C_0(\boldsymbol{p}, \omega) = \frac{2 \tilde \Gamma_0 }{ (\omega + v_0 p_x)^2 + (\Gamma_0 \boldsymbol{p}^2 + m_0)^2}
\end{eqnarray}
A straightforward one loop momentum-shell renormalization group calculation \cite{kardar2007statistical, tauber2014critical, hohenberg1977theory} leads to the following RG recursion relations,
 \begin{eqnarray}
    u_b &=& u_0 b^{4-d}\left[ 1 - \frac{9 u_0}{ 1+ r_0}\ln b \right]
    \label{eq:ub} \\ 
    r_b &=& r_0  b^2\left[ 1 + \frac{3 u_0}{ 1+r_0} \ln b \right] \\ 
    \Gamma_b &=& \Gamma_0 b^{z-2}  
    \label{eq:Gammab}
    \\ 
    v_b &=& v_0 b^{z-1} 
    \label{eq:vb}
\end{eqnarray} 
where $u = \frac{ \tilde \Gamma_0 \lambda_0}{ \Gamma_0^2} K_d \Lambda^{4-d}$ and $r = \frac{m_0 \tilde \Gamma_0}{\Gamma_0} K_d \Lambda^{d-2}$ are respectively the effective coupling and the effective mass of the model, and $z$ is the dynamical exponent.

Intuitively, the reason why the one loop recursion relations coincide with the one of model A is that the same trick that we have used to derive Eq.~(9) in the main text (which is to reabsorb the term proportional to $v_0$ in the Fourier transform of the correlation function)  can be used in all the one loop Feynman diagrams contributing to the renormalization of the parameters.  
To show how this works, let us consider one of the one-loop Feynman diagrams contributing to the renormalization of the coupling constant $\lambda$, 
\begin{equation}
    \Delta \lambda = \quad  18 \times  \includegraphics[valign=c]{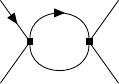}  \quad =  18 \lambda^2 \int_{\Lambda/b} ^ \Lambda \frac{ d^d p}{ (2 \pi)^d }  \int \frac{ d \omega}{ 2 \pi} C_0(p, \omega) G_0( - p, -\omega) = 
\end{equation}

Therefore, we have 
\begin{equation}
    \begin{split} 
    \Delta \lambda &= 18 \lambda^2  \int_{\Lambda/b} ^ \Lambda \frac{ d^d p}{ (2 \pi)^d }  \int \frac{ d \omega}{ 2 \pi} \frac{ 2 \tilde \Gamma_0}{ (\omega+v_0 p_x)^2 +(  \Gamma_0 p^2 + m_0)^2} \frac{1}{ - i (\omega + v_0 p_x) + \Gamma_0 p^2 + m_0} \\
    &= 18 \lambda^2  \int_{\Lambda/b} ^ \Lambda \frac{ d^d p}{ (2 \pi)^d }  \int \frac{ d \omega}{ 2 \pi} \frac{ 2 \tilde \Gamma_0}{\omega^2 +(  \Gamma_0 p^2 + m_0)^2} \frac{1}{ - i \omega + \Gamma_0 p^2 + m_0} \\ 
    & =  \frac{\Omega _d \Lambda^{d-4}}{ ( 2 \pi)^d } \frac{9 \lambda_0 u_0}{ (1 + r_0)^2}  \ln b 
    \end{split} 
\end{equation}
As anticipated the correction $\Delta \lambda$ is independent of the value of the off-equilibrium parameter $v_0$.

The recursion relations of $u_b$, $r_b$ and $\Gamma_b$ are the same ones of model A \cite{hohenberg1977theory}, suggesting that at one-loop level the model belongs to the Ising universality class. As in the equilibrium case, the dynamical exponent $z$ is obtained by setting the scaling dimension of $\Gamma$ to be zero, leading to $ z= 2$. The off-equilibrium parameter $v_0$ does not take any renormalization group correction, but we expect this second result to be a consequence of the one-loop approximations. Furthermore, we expect that a higher order renormalization group calculation could lead to an anisotropic diffusion coefficient $\Gamma \nabla^2 \to \Gamma_{||} \nabla^2 _{||} + \Gamma_{\perp}\nabla_{\perp} ^2$. 

\section{Temporal correlation functions}

The presence of the term $v_0 \de_x \varphi$ in the equation of motion~\eqref{mamal} suggests a non exponential relaxation; the Gaussian correlation function in the $(\bk, t)$ domain reads, 

\begin{eqnarray}
    C_0(\boldsymbol{k}, t)  =  \mathrm{Re} \left ( \frac{\tilde \Gamma_0}{\Gamma_0 \bk^2 + m_0} e^{ + i \omega_{\boldsymbol{k}}t}   \right) 
    \label{eq:Ckt}
\end{eqnarray}
where the frequency $\omega_{\boldsymbol k}$ is defined as, 
\begin{equation}
    \omega_{\boldsymbol k} = i \left( m_0 + \Gamma \boldsymbol{k}^2 \right) + v_0 \boldsymbol{k}_x 
    \label{eq:GaussianOmegaK}
\end{equation}
Close to the critical point, corresponding to $m_0 = 0$, we have
\begin{equation}
    C_0(\boldsymbol{k}, t) = \frac{ \tilde \Gamma_0}{ \Gamma_0} \bk^{-2} e^{- \Gamma_0 k^2 t} \cos(v_0 \bk_x t)
    \label{Capa} 
\end{equation}

\begin{figure}
    \includegraphics[width = 0.4\textwidth]{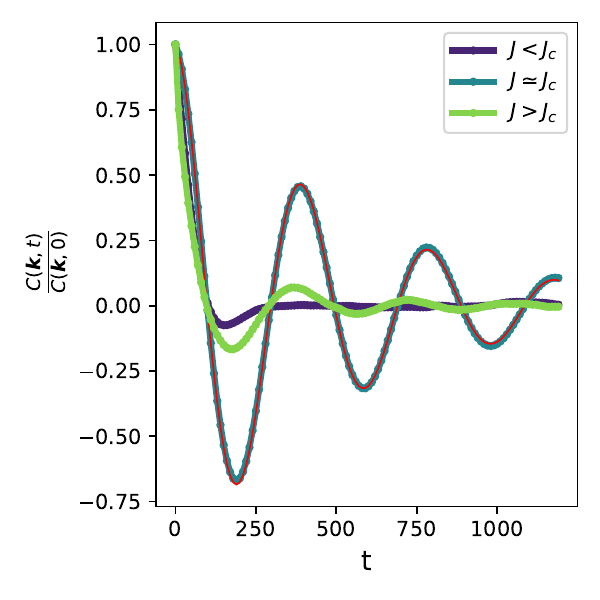}
    \includegraphics[width = 0.4\textwidth]{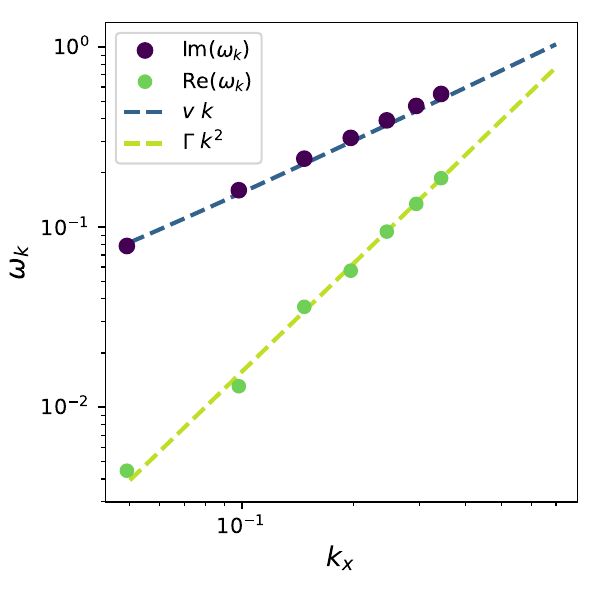}
    \put(-352, 205){\textbf{(a)}}
    \put(-167, 205){\textbf{(b)}}
    \caption{ Temporal correlation function of the off-equilibrium Ising model. \textbf{(a)} The correlation function at $\boldsymbol{k} = 2 \pi /L \times ( 1,0)$ as measured in a numerical simulations of the system. The correlation function is well described by the functional form predicted by the Gaussian theory~\eqref{eq:Ckt}. The red line is a the best fit of~\eqref{eq:Ckt} to the numerical correlation function, where we are fitting the real and the imaginary part of the characteristic frequency $\omega_{\boldsymbol k}$ \textbf{(b)} The real and imaginary part of the characteristic frequency $\omega_{\boldsymbol{ k}}$ as a function of the momentum $\bk$. The frequency is obtained by by fitting Eqs.~\eqref{eq:Ckt} and~\eqref{eq:GaussianOmegaK}. As predicted by the renormalization group calculation the real and imaginary part of $\omega(\boldsymbol k)$ follow the naive scaling ${\rm Im} \omega_{\bk} \simeq k $ and ${\rm Re} \omega_{\bk} \simeq \bk^2$. Data are relative to a system of size $L = 64$ with $\alpha = 0.5$ and $J = 0.99 J_c$.  }
    \label{fig:correlations} 
\end{figure}

To measure the correlation correlation function in the numerical simulations, we first compute the two-dimensional Fourier transform of the configurations $\sigma_{\rm x, y}$, 
\begin{equation}
    \sigma_{\rm k_x, \rm k_y} = \frac{ 1} { \sqrt{N}} \sum_{\rm x,y} \sigma_{\rm x, y} e^{ - i k_x x - i k_y y } 
\end{equation}

and then we compute the correlation function as,
\begin{equation}
    C_{\mathrm{exp}}(\bk,t) = \frac{1}{ T- t} \sum_{t^\prime = 1}^{T - t} \sigma_{\rm k_x, k_y}(t^\prime) \sigma_{\rm k_x, k_y} ^* (t^\prime + t)
\end{equation}
In Fig.~\ref{fig:correlations} we show the correlation function as measured in the numerical simulations, for a system of size $L = 64$ with $\alpha = 0.5$ and $J = 0.99 J_c$. 
We show the best fit to the correlation function of Eq.~\eqref{Capa} - since the correlation function has been normalized the only two parameters we are fitting are the oscillation frequency and the characteristic decay time, which are the real and imaginary part of the characteristic frequency $\omega(\boldsymbol{ k})$.
As we can see in (a) the correlation function is well described by ~\eqref{Capa}. In Fig.~\ref{fig:correlations} (b) we show show the real and imaginary part of the characteristic frequency $\omega(\boldsymbol{ k})$ as function of the momentum $\bk$, again the numerical results are in excellent agreement with the theoretical prediction of Eq.~\eqref{eq:GaussianOmegaK}. It is important to remark that the agreement between the Gaussian scaling observed in Fig.~\ref{fig:correlations} is a prediction of the renormalization group calculation~\eqref{eq:ub}-~\eqref{eq:vb}. In fact having a naive scaling with $\bk$ is a direct consequence  of the fact that the off-equilibrium parameter $v_0$ does not take any renormalization group corrections.

\bibliographystyle{apsrev4-2}
\bibliography{references}